\documentclass[aps,prb,a4paper,twocolumn,showpacs]{revtex4}
\usepackage{graphicx}
\usepackage{epstopdf}
\usepackage{amsmath}
\usepackage{amsfonts}
\usepackage{amssymb}
\usepackage{color}

\makeatletter
\@ifundefined{textcolor}{}
{%
 \definecolor{BLACK}{gray}{0}
 \definecolor{WHITE}{gray}{1}
 \definecolor{RED}{rgb}{1,0,0}
 \definecolor{GREEN}{rgb}{0,1,0}
 \definecolor{BLUE}{rgb}{0,0,1}
 \definecolor{CYAN}{cmyk}{1,0,0,0}
 \definecolor{MAGENTA}{cmyk}{0,1,0,0}
 \definecolor{YELLOW}{cmyk}{0,0,1,0}
 }
\makeatother
\begin{document}
\title{Andreev reflection in a Y-shaped graphene-superconductor device}
\author{Chao Wang, Youlian Zou, Juntao Song$^{\ast}$, Yu-Xian Li$^{\ast}$}
\affiliation{College of Physics $\&$ Information Engineering and Hebei Advanced Thin Films Laboratory,
Hebei Normal University, Shijiazhuang 050024, People's Republic of
China}

\begin{abstract}
Using the non-equilibrium Green function method, we study the Andreev reflection in a Y-shaped graphene-superconductor device by tight-binding model. Considering both the zigzag and armchair terminals, we confirm that the zigzag terminals are the better choice for detecting the Andreev reflection without no external field. Due to scattering from the boundaries of the finite-size centre region, the difference between Andreev retroreflection and specular reflection is hard to be distinguished. Although adjusting the size of the device makes the difference visible, to distinguish them quantitatively is still impossible through the transport conductance. The problem is circumvented when applying a perpendicular magnetic field on the centre region, which makes the incident electrons and the reflected holes propagate along the edge or the interface. In this case, the retroreflected and specular reflected holes from the different bands have opposite effective masses, therefore the moving direction of one is opposite to the other. Which external terminal the reflected holes flow into depends entirely on the kind of the Andreev reflection. Therefore, the specular Andreev reflection can be clearly distinguished from the retroreflected one in the presence of strong magnetic field, even for the device with finite size.

\end{abstract}


\date{\today}

\maketitle

\vskip2pc
\narrowtext
\section{Introduction}

Graphene\cite{Novoselov}, a two-dimensional (2D) allotrope of carbon, was produced by an effective way in recent years. The excitations at low energies are massless, chiral, Dirac fermions\cite{Neto}. Compared to ordinary electrons, Dirac fermions behave in unusual ways in some conditions, such as the anomalous integer quantum Hall effect\cite{Fang,Haldane,Novoselov1} and Klein paradox\cite{Cheianov,Young,Beenakker}. The spectrum of graphene nanoribbons depends on the nature of their edges: zigzag or armchair\cite{Brey}. The zigzag ribbon presents a band of zero-energy modes, which is absent in the armchair case. In previous works, people justified that the zigzag nanoribbons supports current carrying zero-modes localized along the edges\cite{Nakada}. When a nonuniform magnetic field is applied perpendicular to a graphene ribbon, snake states\cite{Oroszlany} are formed and expected to be resilient against scattering by impurities at edge states. Then for a uniform magnetic field in a graphene \emph{p-n} junction\cite{Williams}, snake states travel along the \emph{p-n} interface due to a sign change of charge carrier's effective mass across the junction reversing the direction of the Lorentz force. The understanding and control of the electron travelling in the graphene can open doors for a new frontier in electronics.

The Andreev reflection\cite{Andreev} is an important transport process between a conductor and a superconductor that an electron incident from the conductor is reflected as a hole, while two electrons enter the superconductor as Cooper pair. When the bias is smaller than the superconductor gap, the conductance of the metal-superconductor hybrid device is mainly determined by the Andreev reflection. For a normal conductor, the Andreev reflection is retroreflection in which the path of the reflected hole is back along the path of the incident electron. The crossed Andreev reflection also belongs to conventional Andreev reflection. It is that the incident electron enters from one side of the superconductor and the reflected hole flows out from the other side of the superconductor if the centre superconductor is not very wide. Recently, in some conditions the crossed Andreev reflection can take place when the width of the superconductor is as large as the superconducting coherent length\cite{Hou,Deng}.  For the graphene-superconductor system, different from the conventional retroreflction, the specular Andreev reflection in which the hole is reflected along the specular direction of the incident electron is an intriguing phenomena proposed by Beenakker\cite{Beenakker}.

\begin{figure}[]
\centering
\includegraphics[width=1.0\columnwidth,angle=0]{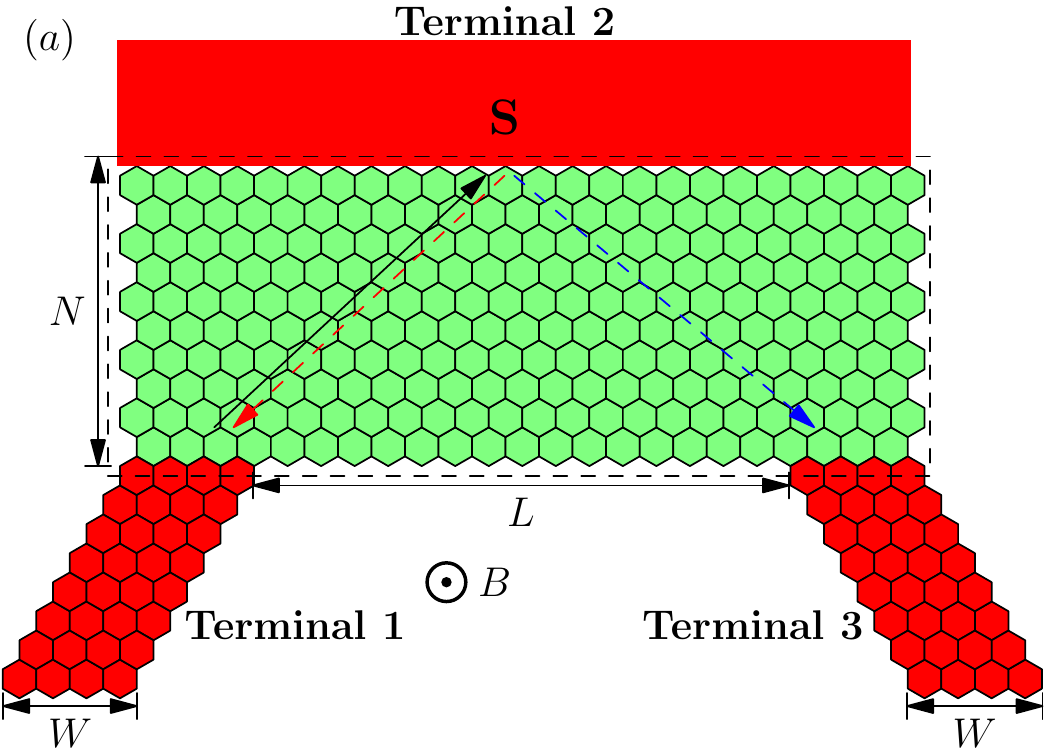}\\[10pt]
\centering
\includegraphics[width=1.0\columnwidth,angle=0]{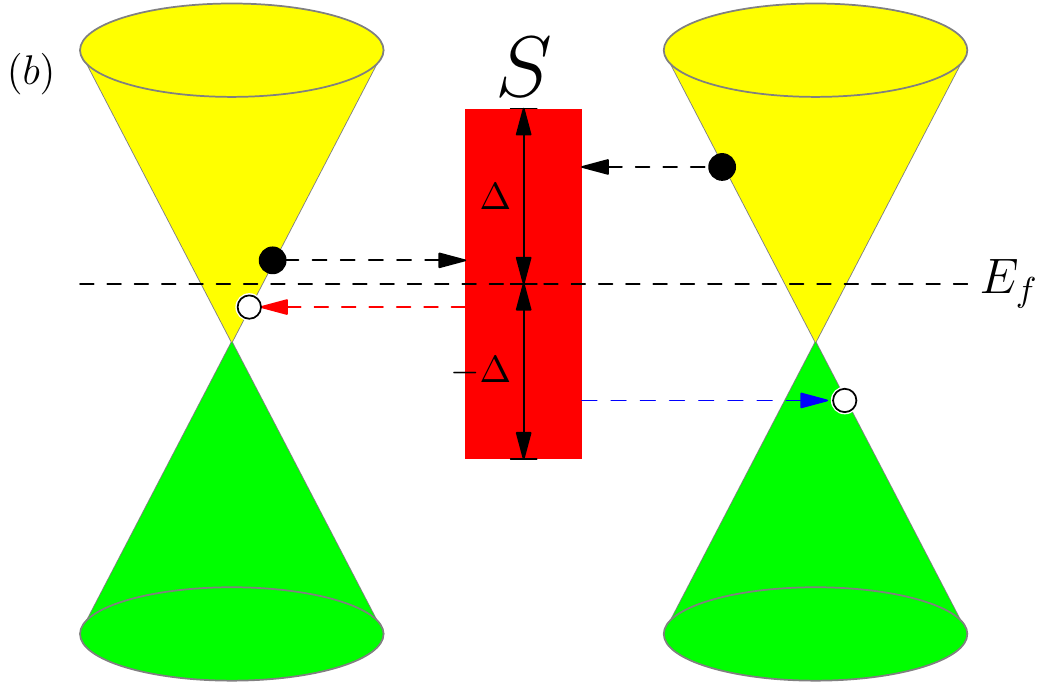}\\
\caption{(Color online) (a) The schematic diagram for a Y-shaped graphene-superconductor device under a perpendicular magnetic field. In this diagram, the widths of the centre region and the graphene terminals are $N=5$ and $W=4$, respectively. The distance between terminal $1$ and $3$ is $L=15$ showed in this figure. When the incident electrons (black solid line) come from the terminal 1, in the case of $B=0$ the retroreflected holes (red dashed line) and the specular reflected holes (blue dashed line) should flow into terminal 1 and 3, respectively. (b) The schematic diagram for Andreev retroreflection (left) and specular Andreev reflection (right). }\label{fig1}
\end{figure}

Ever since, many research groups have studied the graphene-suprcondutor hybrid system\cite{Titov,Xie,Efetov,Linder,Cheng,Xing,Rainis}. From the physical point of view, if the incident electron and the reflected hole are from the same band (the conduction or valence band), the conventional retroreflection (the intraband Andreev reflection) will happen. In contrast, the specular Andreev reflection (the interband Andreev reflection) occurs when the incident electron and the reflected hole come from the conduction band and the valence band, respectively. For the normal metal-superconductor hybrid system, there is only intraband Andreev reflection because the excitation gap between the conduction and valance bands is bigger than the superconducting gap of the superconductor. However, the Fermi energy is generally close to Dirac point for the graphene [see Fig. \ref{fig1}(b)], so interband Andreev reflection, namely specular Andreev reflection, can take place in the graphene-suprcondutor hybrid system. Due to time-reversal symmetry, electron excitations in one valley($K$) are related to hole excitations in the other valley($K'$). In this sense, both the specular reflection and retroreflection occur in a two-terminal graphene-superconductor device. Then the question arises: How to distinguish the specular Andreev reflection from the conventional retroreflection in experiment. Many groups have studied this question in different kinds of graphene-supercondutor hybrid systems\cite{Cheng,Xing,Rainis,Schelter}. It was reported that in a four-terminal graphene-superconductor hybrid systems\cite{Cheng}, the retroflection and specular reflection can be controlled and separated by controlling the phase of the supercondutor terminals. In a graphene-based ferromagnet/superconductor junction, the ferromagnetic exchange interaction\cite{Xing}, which can enhance specular Andreev reflection but suppress Andreev retroreflection, can be used to detect the specular Andreev reflection.

In this paper, we propose a Y-shaped device with finite size [see Fig. \ref{fig1} (a)] to study the Andreev reflection, and consider the graphene terminals as the zigzag ribbons and the armchair ribbons. Firstly, the Andreev reflection is studied in this device without a magnetic field. If the incident electrons come from the terminal 1, at the graphene-superconductor interface the retroreflected holes and the specular reflected holes should be reflected and flow into the terminal 1 and the terminal 3, respectively. It should be pointed that the centre region used in this work is a finite-size monolayer graphene film  which is more like a real device in experiment. Because of the intervalley scattering in the two-dimensional materials\cite{Zhou}, both kinds of the reflected holes can flow into either of the graphene terminals. The Andreev reflection coefficients of the different terminals are compared, and the zigzag terminal is a better choice for detecting the Andreev reflection. Different from the pervious work\cite{Cheng}, it is found in this work that the specular Andreev reflection can be distinguished from the conventional Andreev reflection (retroreflection) in this Y-shape device, but the difference is not obvious due to the scattering effect from edges.

After applying an uniform magnetic field perpendicular to the device, however, the incident electrons will travel along the edge of the device\cite{Chen}, which makes the scattering negligible. On the interface between the graphene and superconductor, the Andreev reflected electrons and holes form the cyclotron orbits\cite{Akhmerov} along the boundary. Due to the retroreflected holes and the specular reflected holes coming from the different bands, the helical direction of them is contrary to each other\cite{Rickhaus}. So when the incident electrons come from the terminal 1, the retroreflected holes flow into terminal 3 while the specular reflected holes are reflected backwards to the terminal 1 in a suitable magnetic field. Experimentally, through adjusting the incident energy, the current, flowing into terminal 1 or terminal 3, can be detected to distinguish the Andreev reflection in a magnetic field strong enough.

The rest of this paper is arranged as follows. In Sec.\uppercase\expandafter {\romannumeral 2}, the model Hamiltonian for the system is presented and the formulas for calculating the current and the Andreev reflection coefficients are derived. Our main results are shown and discussed in Sec.\uppercase \expandafter {\romannumeral 3}. Finally, a brief conclusion is presented in Sec.\uppercase\expandafter {\romannumeral 4}.

\section{Model and Formula}

The graphene-superconductor device  proposed here is a Y-shaped model which consists of two graphene terminals (terminal 1 and 3) and a superconductor terminal (terminal 2) connected  to the central graphene region [see Fig.\ref{fig1} (a)]. By using the nonequilibrium Green function method, the Andreev reflection coefficients are calculated. The Hamiltonian of the device is
\begin{equation}\
 H=H_{C}+H_{S}+H_{1}+H_{3}+H_{T},
\end{equation}
where $H_{C}$, $H_{S}$, $H_{1/3}$ and $H_{T}$ are the Hamiltonians of the centre region, superconductor terminal 2, graphene terminals 1 and 3, coupling between the graphene and superconductor terminal 2, respectively.

In the tight-binding model, $H_{C}$ and $H_{1,3}$ are given by
\begin{eqnarray}
H_{C}/H_{1/3}=\sum_{i}E_{0}a^{\dag}_{\emph{i}}a_{\emph{i}}-\sum_{\langle\emph{ij}\rangle}te^{i\phi_{ij}}a_{\emph{i}}^{\dagger}a_{\emph{j}},
\end{eqnarray}
where $a_{\emph{i}}^{\dagger}$ ($a_{\emph{i}}$) is the creation (annihilation) operator at site \emph{i} and $E_{0}$ is the energy
for Dirac points. The second term in the Hamiltonian stands for the nearest-neighbor hopping. The magnetic factor $e^{i\phi_{ij}}$ comes from Peierls substitution and $\phi_{ij}=\int_{i}^{j}\textbf{A}\cdot\emph{d\textbf{l}}/\Phi_{0}$ where  $\textbf{A}=(-By,0,0)$ is magnetic vector potential and $\Phi_{0}=\hbar/e$. The magnetic field $B$ is applied to the centre region along the perpendicular direction. We consider that the graphene region is directly coupled to a {\sl s}-wave surperconductor terminal. Described by a continuum model, the superconductor terminal is represented by BCS Hamiltonian,
\begin{equation}
H_{\emph{S}}=\sum_{\textbf{\emph{k}},\sigma}\varepsilon_{\textbf{\emph{k}}}C^{\dagger}_{\emph{\textbf{k}}\sigma}C_{\textbf{\emph{k}}\sigma}
+\sum_{\textbf{\emph{k}}}(\Delta^{}C_{\textbf{\emph{k}}\downarrow}C_{-\textbf{\emph{k}}\uparrow}
+\Delta^{*}C^{\dagger}_{-\textbf{\emph{k}}\uparrow}C^{\dagger}_{\textbf{\emph{k}}\downarrow}),
\end{equation}
where $\Delta=\Delta_{\emph{0}}e^{\emph{i}\theta}$. Here $\Delta_{\emph{0}}$ is the superconductor gap and $\theta$ is the superconductor phase. The coupling between superconductor terminal and graphene is described by
\begin{equation}
H_{\emph{T}}=\sum_{\emph{i},\sigma}\emph{t}a^{\dagger}_{\emph{i}\sigma}C_{\sigma}(x_{\emph{i}})+h.c.
\end{equation}
Here $x_{\emph{i}}$ is the horizontal position of the carbon atom $\emph{i}$ and $C_{\sigma}(x)=\sum_{\emph{k}_{x},\emph{k}_{y}}e^{\texttt{i}k_{x}x}C_{\textbf{\emph{k}}\sigma}$.

The current through the terminal 1 can be obtained by using the Heisenberg equation\cite{Cheng} of motion:
\begin{equation}
\begin{split}
\emph{I}_{1}= &\frac{2e}{\hbar}\int\frac{dE}{2\pi}[(\textit{f}_{1+}-\textit{f}_{2})T_{12}+(\textit{f}_{1+}-\textit{f}_{1-})T_{11A}\\&+(\textit{f}_{1+} -\textit{f}_{3-})T_{13A}+(\textit{f}_{1+}-\textit{f}_{3+})T_{13}],
\end{split}
\end{equation}
where $\textit{f}_{\alpha\pm}=1/\{\exp[(E \mp eV_{\alpha})/k_{B}T]+1\}$ and
$\textit{f}_{2}=1/\{\exp(E/k_{B}T)+1\}$ are the Fermi distribution function of three terminals. Here we set the bias of the superconductor terminal at zero. Noted that, $T_{12}=Tr\{\Gamma_{1\uparrow\uparrow}[G^{r}\Gamma_{2}G^{a}]_{\uparrow\uparrow}\}$ and $T_{13}=Tr\{\Gamma_{1\uparrow\uparrow}G^{r}_{\uparrow\uparrow}\Gamma_{3\uparrow\uparrow}G^{a}_{\uparrow\uparrow}\}$
are the normal transmission coefficients from terminal 1 to terminal 2 and 3, respectively. $T_{1nA}=Tr\{\Gamma_{1\uparrow\uparrow}G^{r}_{\uparrow\downarrow}\Gamma_{n\downarrow\downarrow}G^{a}_{\uparrow\downarrow}\}$ is the Andreev reflection coefficients for the incident electron coming from the terminal 1 with the hole reflected to the terminal 1 or 3. The subscripts $\uparrow\uparrow$, $\downarrow\downarrow$, $\uparrow\downarrow,$ and $\downarrow\uparrow$ represent the 11, 22, 12, and 21 matrix blocks in the function. $\Gamma_{\alpha}(E)=i[\mathbf{\Sigma}_{\alpha}^{r}-(\mathbf{\Sigma}_{\alpha}^{r})^{\dagger}]$
is the linewidth function and $G^{r(a)}$ is the retarded (advanced) Green function of the central region in Nambu represention. $\mathbf{\Sigma}_{\alpha}^{r}$ is the retarded self-energy
due to the coupling to the terminal $\alpha$. $\mathbf{\Sigma}_{\alpha,ij}^{r}(E)=tg_{\alpha,ij}^{r}(E)t$,  where $g_{\alpha,ij}^{r}(E)$  is the surface Green function of the terminal $\alpha$. We can numerically calculate the surface Green function of the graphene terminals 1 and 3. For the superconductor terminal 2, the surface Green function is
\begin{equation}\label{1}
 g_{2,ij}^{r}(E)=i\pi\rho\beta(E)J_{0}[k_{f}(x_{i}-x_{j})]\left(
\begin{array}{cc}
  1 & \Delta/E \\
  \Delta^{\ast}/E & 1 \\
\end{array}
\right),
\end{equation}
where $\rho$ is the normal density of states. $J_{0}[k_{f}(x_{i}-x_{j})]$ is the Bessel function of the first kind with the Fermi wavevector $k_{f}$, and $\beta(E)=-iE/\sqrt{\Delta_{0}^{2}-E^2}$ for $|E|<\Delta_{0}$ and $\beta(E)=|E|/\sqrt{E^2-\Delta_{0}^{2}}$ for  $|E|>\Delta_{0}$. Then we can calculate the Green function $G^{r}(E)=1/(EI-H_{C}-\sum_{\alpha=1,2,3}\mathbf{\Sigma}^{r}_{\alpha})$, where $I$ is the unit matrix and  $H_{C}$ is the Hamiltonian of the center region in the Nambu representation.

\section{Results and discussion}
In the numerical calculation, we take the hopping energy $t=2.75$ $eV$, and the length of the nearest-neighbor C-C bond $a_{0}=0.142$ $\mathrm{nm}$ as in the real graphene sample. The widths of the graphene terminals 1 and 3 are set to $W=50$. The superconductor gap and phase are set to $\Delta_{0}=t/2750=1$ $\mathrm{meV}$ and $\theta=0$, respectively. The Fermi wavevector is $k_{f}=10$ $\mathrm{nm^{-1}}$. It should be pointed out especially that the Fermi energy is always fixed at $E_{f}=0$ in the whole work, which consequently leads the Dirac points to be at $E_0$.

\begin{figure}[tbp]
\centering
\includegraphics[width=1.0\columnwidth,angle=0]{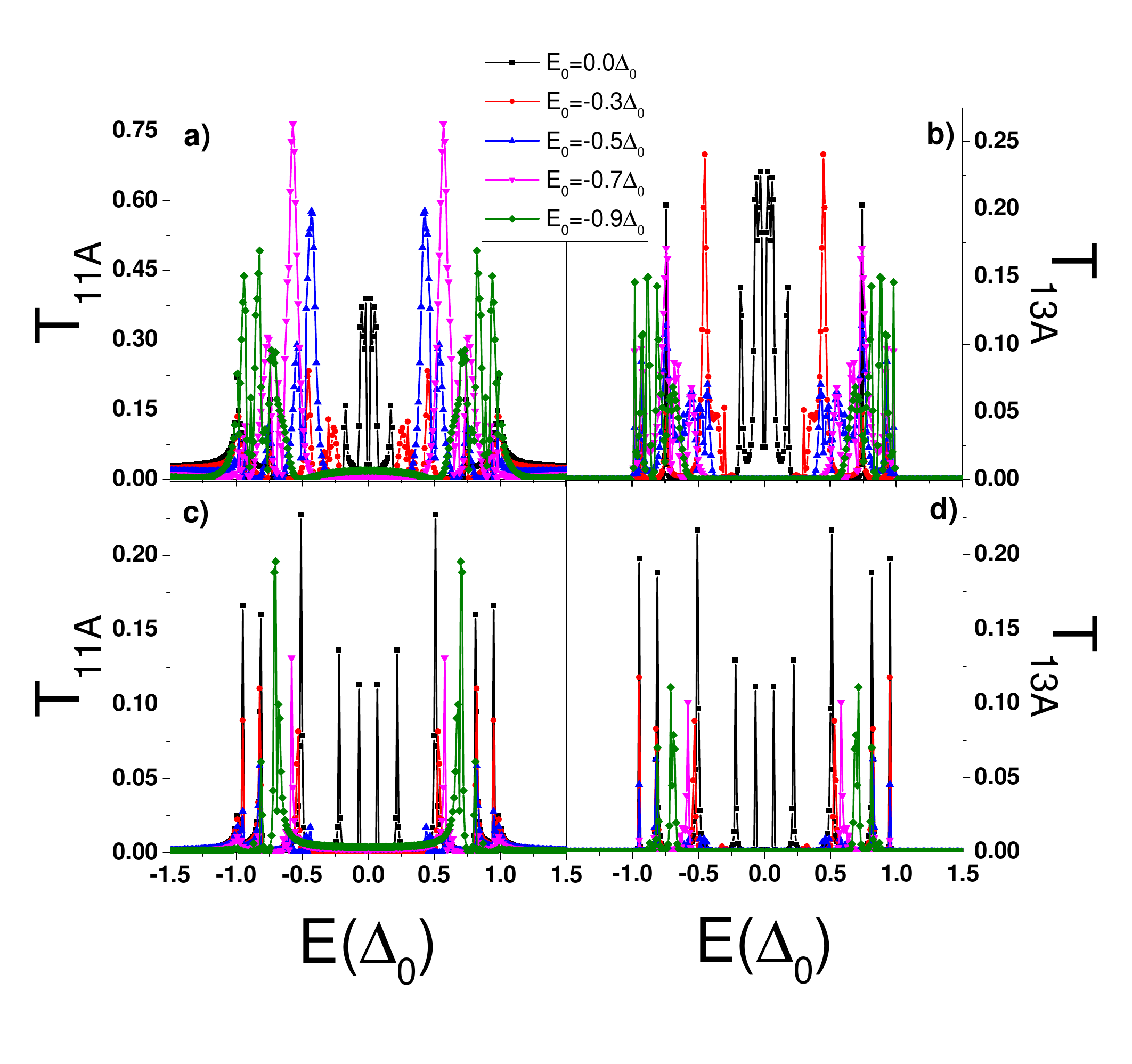}
\caption{(Color online) $T_{11A}$ and  $T_{13A}$ vs the incident energy $E$ for $N=50$ and $L=300$ when the terminal 1 and 3 are the zigzag ribbon [a) and b)] or the armchair ribbon [(c) and (d)].} \label{fig2}
\end{figure}

Fig. \ref{fig2} shows the Andreev reflection coefficients $T_{11A}$ and  $T_{13A}$ as a function of the incident energy $E$. Firstly, the graphene terminals connected with the centre region are set to be the zigzag ribbons [see Fig.\ref{fig2} (a) and (b)]. For $|E| < \Delta_0$, the curves in Fig.\ref{fig2} (a) look like complicated at first sight. However, it is very simple once we notice the fact that the dominating peaks in Fig.\ref{fig2} (a) appear at the energies of Dirac points $E=\pm E_{0}$ or of the superconductor gap edge $E=\pm\Delta_{0}$. For example, there are four peaks at $E=\pm 0.5\Delta_{0}$ and $E=\pm \Delta_{0}$ for the blue line with $E_0=-0.5\Delta_{0}$. In general, some side peaks (or dips) will emerge nearby the peak of $E=\pm \Delta_{0}$, but their heights are comparatively small. For $|E| > \Delta_0$, the $T_{11A}$ decays quickly to zero as the incident energy $E$ is tuned away from the superconductor gap.

Although the peaks of $T_{13A}$ decrease remarkably in Fig.\ref{fig2} (b), all curves of $T_{13A}$ have very similar shapes as those of $T_{11A}$. Given the Fermi energy has been fixed at $E_f=0$, a basic rule can be easy to deduce that there are no specular Andreev reflections for $|E|<|E_0|$, and in contrast no Andreev retroreflections for $|E|>|E_0|$. However, the curves in Fig.\ref{fig2} (a) and (b) violate the above rule. Let us still take the curve of $E_0=-0.5\Delta_{0}$ as example. Note that, high peaks up to $0.58$ appear nearby $E=\pm 0.48\Delta_0$ in Fig.\ref{fig2} (a), which should be attributed to the Andreev retroreflection, while the peaks at $E=\pm 0.52\Delta_0$ also arise, which should not be there according to the reflection rule above. When we turn to the corresponding case in Fig.\ref{fig2} (b), $E_0=-0.5\Delta_{0}$, there are many small peaks nearby $E=\pm 0.48\Delta_0$ and a comparatively big peak at $E\approx\pm 0.75\Delta_0$. The reflection rule mentioned above is also ruined in the curves of $T_{13A}$.

While choosing the graphene terminals to be armchair ribbons in Fig.\ref{fig2} (c)  and (d), there are more sharp peaks than those in Fig.\ref{fig2} (a)  and (b). Except these sharp peaks, Andreev reflection coefficients are almost be zero in Fig.\ref{fig2} (c)  and (d). Comparing the plots in Fig.\ref{fig2} (c) and (d), it is found that there is no qualitative difference between them: the positions of and the heights of peaks are almost the same. It means that the reflected holes can flow into both the armchair graphene terminal 1 and 3, no matter what kind of the Andreev reflection is. The reason is that the intervalley scattering\cite{Nakada} for the armchair terminals is more strong than that for the zigzag terminals, which makes the zigzag terminals present better performances than the armchair terminals. In conclusion, we can not distinguish the Andreev retroreflection from the specular one by using armchair ribbons too.

From the curves in Fig.\ref{fig2}, it becomes rather difficult to distinguish the retroreflection process and the specular reflection process accurately in experiment, even though the heights of peaks have changed somewhat when different reflections dominates in \ref{fig2} (a) and (b). It has been intensively reported that the Andreev retroreflection can be well distinguished from the specular one by using a continuum model\cite{Beenakker,Xing}. The finite size adopted in this work could give rise to scattering from the edges of centre region and the intervalley, which ruin the ideal rule which the Andreev reflections should obey. In this sense, it is very necessary to find out how the size of centre graphene affects the Andreev reflection.

\begin{figure}[tbp]
\includegraphics[width=1.0\columnwidth,angle=0]{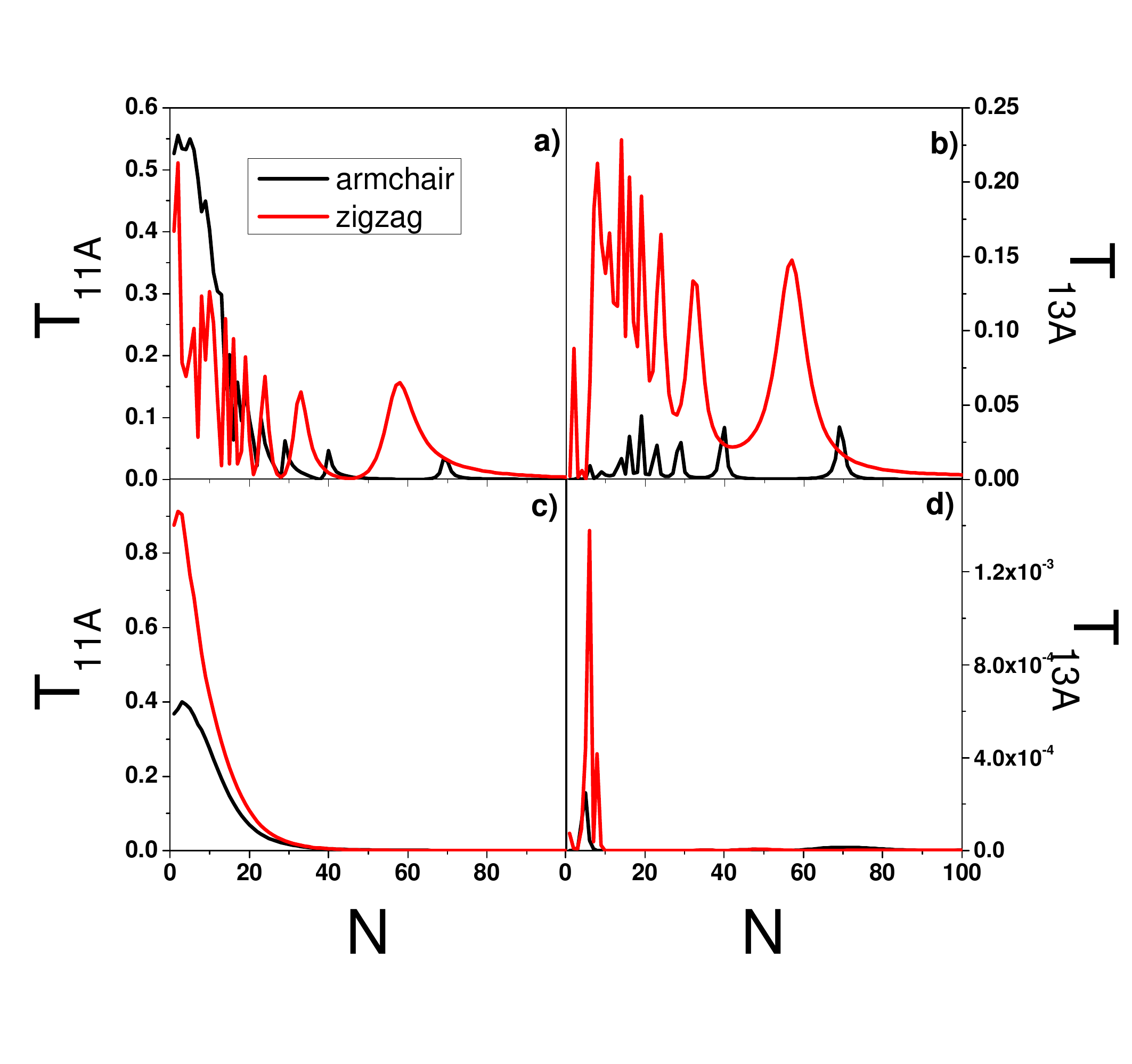}
\caption{(Color online) $T_{11A}$ and  $T_{13A}$ vs the width $N$ with $E_{0}=-0.5\Delta_{0}$. The incident energy are $E=0.7\Delta_{0}$ [a) and b)] and $E=0.01\Delta_{0}$ [c) and d)], respectively. The distance between the two graphene terminals is $L=300$. } \label{fig3}
\end{figure}

In Fig. \ref{fig3}, we show the behavior of Andreev reflection coefficients\cite{Xie,Cheng} as changing the width of the device. For the sake of simplification in following discussion, we adopted a fixed value of $E_{0}=-0.5\Delta_{0}$ in Fig. \ref{fig3}. When the incident energy $|E|$ is smaller than that of Dirac points, $|E_{0}|$,  only the Andreev retroreflection should happen. This point is confirmed in Fig.\ref{fig3} (c) and (d) with $E=0.01\Delta_{0}$. Noted that, $T_{13A}$ is as small as ten to the power of minus three in Fig.\ref{fig3} (d)  while in Fig.\ref{fig3} (c) $T_{11A}$ shows large transmission coefficient for the width not greater than $N=20$. It can be seen that the maximum of $T_{11A}$ can reach up to nearly $0.9$ and $0.4$ for zigzag and armchair ribbons, respectively. However, the $T_{11A}$ decays quickly to zero when the width $N$ continues growing and surpasses the threshold value $40$. These findings can be also confirmed in Fig.\ref{fig2}.

According to the theory proposed by Beenakker\cite{Beenakker}, it is anticipated to observe only the specular reflection process when the incident energy $|E|$ is greater than that of Dirac points $|E_{0}|$. However, in Fig.\ref{fig3} (a) and (b) with $E=0.7\Delta_{0}$, both $T_{11A}$ and $T_{13A}$ show oscillating behavior with almost similar variation. It is noted that the transmission coefficients oscillate rapidly at smaller width, show distinct peaks one after another from $N=20$ to $60$, and then decays quickly to zero for the width greater than $N=60$. Even though $T_{11A}$ shows very small values at the dips, especially at $N=45$, and in this case $T_{13A}$ shows nonzeros value, $T_{13A}\approx 0.05$, this condition is almost uncontrollable in experiment. To search a suitable condition is almost an impossible task.

From Fig.\ref{fig3}, the behaviors of transmission coefficients have showed clearly that the width of centre region affects the reflection processes. Firstly, the quick decay of transmission coefficients as increasing the width indicates that an appropriate width should be chosen so as to focus all the reflected holes flowing into the terminal 1 or 3. Secondly, the oscillating behaviors of transmission coefficients demonstrates that the scattering from edges and intervalley has also play an important role in observing the specular Andreev reflection.

\begin{figure}[tpb]
\includegraphics[width=1.0\columnwidth,angle=0]{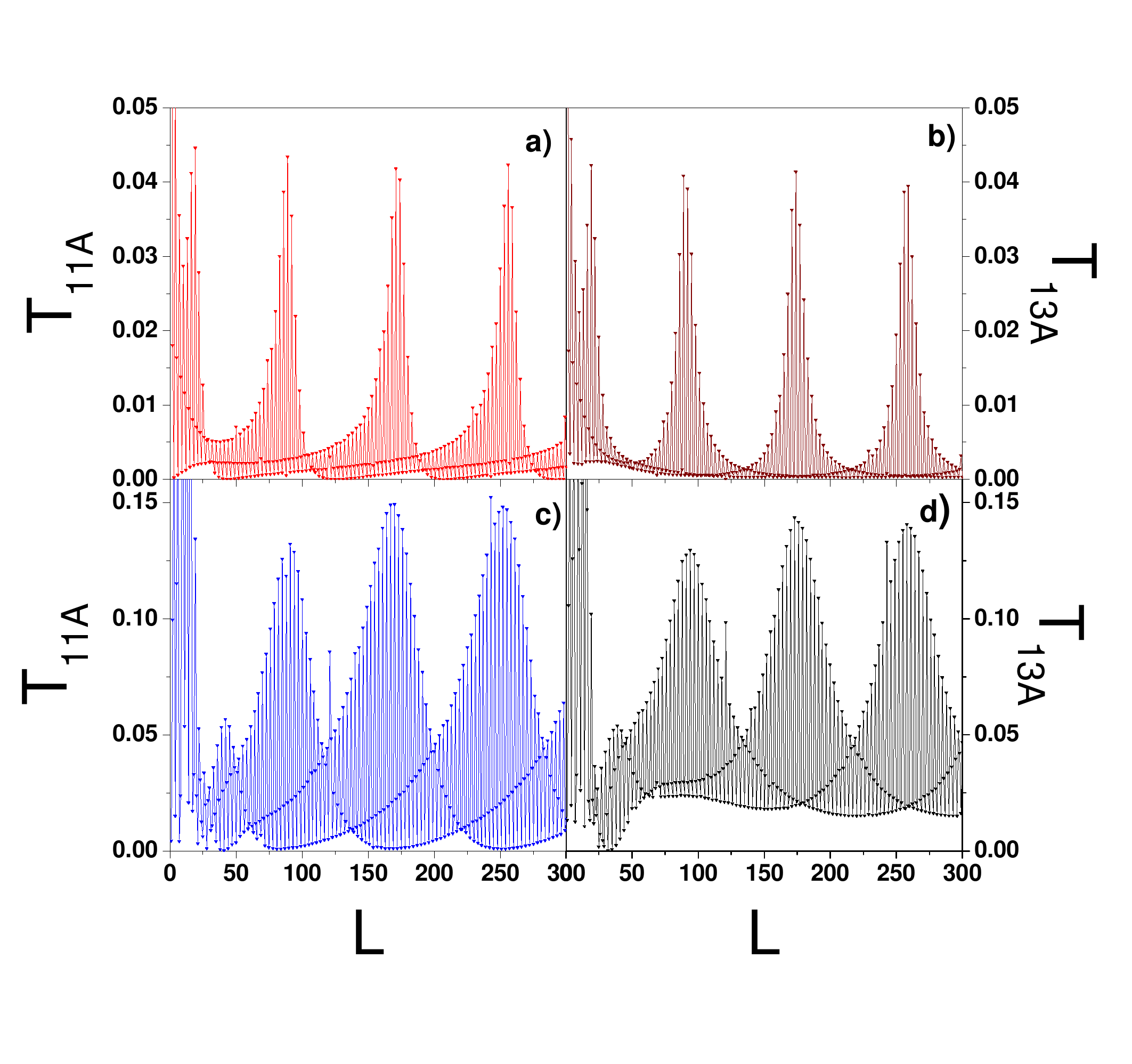}
\caption{(Color online) $T_{11A}$ and  $T_{13A}$ vs the distance $L$ with $E_{0}=-0.5\Delta_{0}$ and $E=0.7\Delta_{0}$ when the terminal 1 and 3 are the armchair ribbon [(a) and (b)] or the zigzag ribbon [(c) and (d)].  The width of the device is $N=50$. } \label{fig4}
\end{figure}

What we are interested in is whether two kinds of Andreev reflection can be distinguished in this device. Now, we have seen that the width of the device has affected Andreev reflection coefficients. In Fig. \ref{fig3}, we see that there are $T_{11A}=0$ and $T_{13A}\neq0$ for $|E|>|E_{0}|$ when the width of the device is exactly the appropriate value. We want to know how the Andreev reflection coefficients change by adjusting the distance between the two graphene terminals. Detecting the specular Andreev reflection is one of our research focuses, so we set $E=0.7\Delta_{0}$ and $E_{0}=-0.5\Delta_{0}$ in Fig. \ref{fig4}.

By adopting a fixed width $N=50$, we plot $T_{11A}$ and  $T_{13A}$ versus the distance between the terminal $1$ and terminal $3$ in Fig. \ref{fig4}. We can see that $T_{11A}$ and  $T_{13A}$ show the oscillation with the distance increasing no matter what the terminals are.  When the distance is more than $50$, the oscillation of the Andreev reflection coefficients is periodic as the distance increases. Although Andreev retroreflection should be banned and only specular Andreev reflection is permitted for the parameters adopted in Fig. \ref{fig4}, $T_{11A}$, corresponding to Andreev retroreflection, shows almost similar behaviors as $T_{13A}$, corresponding to specular Andreev reflection. Combining the oscillating behaviors of $T_{11A}$ and  $T_{13A}$, it can be concluded that scattering from edge and interface ruins the ideal process of specular reflection and retroreflection.

Note that there are some difference between $T_{11A}$ and  $T_{13A}$ in Fig. \ref{fig4}. For example, the minimum value of $T_{11A}$ is zero with changing the distance in Fig.\ref{fig4} (c), and the minimum value of $T_{13A}$ keeps always nonzero for $L>50$ for zigzag terminals in Fig.\ref{fig4} (d). However, these weak distinctions are hard to distinguish in experiment once much more noise is inevitably included. As is known, electrons and reflected holes would move forward in their separate ways in presence of magnetic field. In recent, the different trajectories of reflected hole is experimentally observed in Josephson junction\cite{Amet2016} and be confirmed in theory\cite{Liu2017}. Thus, an external magnetic field should be a very effective tool to distinguish the holes in specular reflection and retroreflection.

\begin{figure}[tbp]
\includegraphics[width=1.0\columnwidth,angle=0]{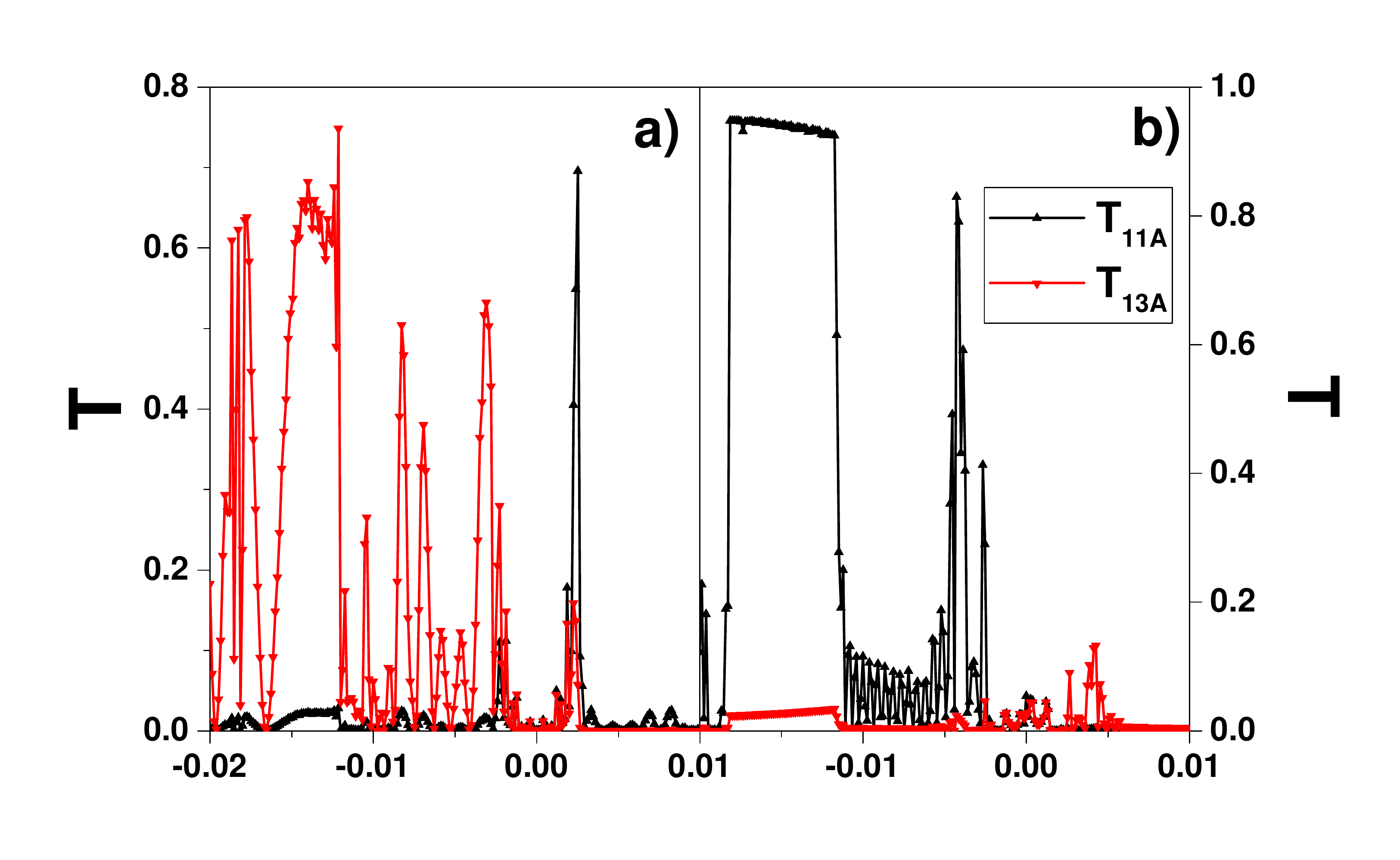}
\caption{(Color online) $T_{11A}$ and  $T_{13A}$ vs the magnetic field $\phi$ with $E_{0}=-0.5\Delta_{0}$ and $N=50$. The incident energies are $E=0.1\Delta_{0}$ in (a) and $E=0.95\Delta_{0}$ in (b), respectively.} \label{fig5}
\end{figure}

In Fig.\ref{fig5}, we plot the Andreev reflection coefficients with changing the magnetic field. When the magnetic field is applied and $\phi<0$, the incident electrons from terminal 1 travel to the interface along the left edge of the device and which terminal the reflected holes flow into depends on the kind of the Andreev reflection. Because the cyclotron radius of the electrons and the holes is too large to form the edge states\cite{Chen} in presence of weak magnetic field, the reflection coefficients oscillate clearly with the magnetic field changing from $0$ to $-0.01$. In Fig. \ref{fig5} (a) with $E_{0}=-0.5\Delta_{0}$ and $E=0.1\Delta_{0}$, when the magnetic field is changed from $-0.01$ to $-0.015$, $T_{13A}$ reaches up to more than $0.6$ while $T_{11A}$ is close to zero. When setting $E_{0}=-0.5\Delta_{0}$ and $E=0.9\Delta_{0}$ in Fig. \ref{fig5} (b), $T_{11A}$ maintains stable and reaches up to more than $0.9$ as changing the magnetic field $\phi$ from $-0.01$ to $-0.02$.  To sum up, $T_{11A}$ and $T_{13A}$ exhibit two distinctive behaviors in two different cases, as shown in Fig. \ref{fig5} (a) and (b), and now it becomes feasible to distinguish Andreev retroreflection and specular Andreev reflection.

In the absence of magnetic field, $T_{11A}$ should give a non zero value while $T_{13A}$ should show a zero value for $E_{0}=-0.5\Delta_{0}$ and $E=0.1\Delta_{0}$, where ideally only Andreev retroreflections happen; they reversed roles in the case of $E_{0}=-0.5\Delta_{0}$ and $E=0.95\Delta_{0}$, where specular Andreev reflections are permitted. However, one may notice that the behaviors of $T_{11A}$ and $T_{13A}$ in Fig. \ref{fig5} is totally contrary to the rule above. It is because that separate Landau levels as chiral states will come into being when external magnetic field is strong enough.

In the presence of strong magnetic field, the direction of reflected hole in Andreev retroreflections would be bent along the same direction of the incident electron from the terminal 1, and the reflected hole would be reflected  again and re-transformed into an electron due to Andreev reflection. This two kinds of reflection processes alternating till an electron moves into the terminal 3. Namely, composite edge states consisting of incident electrons and reflection holes emerge near the interface between graphene and superconductor, which helps an incident electron from the terminal 1 tunneling into the terminal 3. In this sense, Andreev retroreflections with magnetic field would be measured by nonzero transmission coefficient $T_{13A}$ between the terminals 1 and 3. On the other hand, the direction of reflected holes in specular Andreev reflection would be bent into the reverse direction of the incident electron, and thus the reflected hole, which cannot go ahead along the interface between graphene and superconductor, would finally tunnel into the terminal 1. This means that the process of specular Andreev reflection gives a nonzero transmission coefficient $T_{11A}$ in the presence of magnetic field. How the reflected hole in two Andreev reflections moves in the presence of a suitable magnetic field and how the composite chiral edge states forms have been discussed in more detail in Refs. [\onlinecite{Amet2016}] and [\onlinecite{Liu2017}].

When turning to Fig.\ref{fig5} again, it is not difficult to understand the behaviours of $T_{11A}$ an $T_{11A}$. Within the range from $\phi=-0.017$ to $-0.012$, for $|E|<|E_{0}|$ and $|E|>|E_{0}|$ the difference between $T_{11A}$ and $T_{13A}$ is obvious. That is to say we can distinguish the two kinds of Andreev reflections clearly in this device through controlling the magnetic field. When the magnetic field changes its direction ($\phi>0$), the electrons injected from terminal 1 flow directly into the terminal 3 due to the reverse Lorentz force and chiral states,  no Andreev reflection happens for an incident electron from the terminal 1. Therefore, both $T_{11A}$ and $T_{13A}$ are close to zero when external magnetic field with $\phi>0$ is strong enough.

\section{Conclusions}
In our work, we study the Andreev reflection in a Y-shaped graphene-superconductor device. Different from previous works\cite{Beenakker,Xing}, the intervalley scattering effect and the boundary effect are naturally included in the real device used in our calculation. Due to the intervalley scattering, the reflected holes can flow out from both of the graphene terminals, no matter what kind of the Andreev reflection is. It is showed that the zigzag terminals are the better choice to detect the Andreev reflection in the graphene-superconductor hybrid system.

The Andreev reflection coefficients are also related to the size of the device. For $|E|<<|E_{0}|$, $T_{11A}$ is close to zero when the width of the device is large enough. Adjusting the position of the probe, we can get $T_{11A}=0$ and $T_{13A}\neq0$ for $|E|>|E_{0}|$ with the certain width of the device. That means that the specular Andreev reflection can be distinguished in the Y-shaped graphene-superconductor device. But the difference is not obvious due to the strong scattering effect. By applying a perpendicular magnetic field on the device to weaken the scatting effect, the specular Andreev reflection can be distinguished from retroreflection through controlling the magnetic field. Our research findings suggest that the Andreev reflections can be distinguished experimentally in the presence of appropriate magnetic field.

\section*{ACKNOWLEDGMENTS}

This work was supported by the National Natural Science Foundation of China (Grant No. 11474085), the Natural Science Foundation of Hebei (Grant No. A2017205108), the youth talent support program of Hebei education department (Grant No. BJ2014038), the Outstanding Youth Foundation of HBTU (Grant No. L2016J01), and the youth talent support program of Hebei Province.


\end{document}